\documentclass[prl,aps,twocolumn,superscriptaddress,showpacs]{revtex4-1}

\usepackage[dvips]{graphicx} % for figures
\usepackage{amsfonts}
\usepackage{amssymb}
\usepackage{amscd}
\usepackage{amsmath}    % need for subequations
\usepackage{enumerate}
\usepackage{epsfig}
\usepackage{subfigure}

\begin{document}

\title{Field Test of Measurement-Device-Independent Quantum Key Distribution}

\author{Yan-Lin Tang}
\thanks{These authors contributed equally to this work}
\author{Hua-Lei Yin}
\thanks{These authors contributed equally to this work}
\affiliation{Department of Modern Physics and National Laboratory for Physical Sciences at Microscale, Shanghai Branch, University of Science and Technology of China, Hefei, Anhui 230026, China}
\affiliation{CAS Center for Excellence and Synergetic Innovation Center in Quantum Information and Quantum Physics, Shanghai Branch,  University of Science and Technology of China, Hefei, Anhui 230026, China}
\author{Si-Jing Chen}
\thanks{These authors contributed equally to this work}
\affiliation{State Key Laboratory of Functional Materials for Informatics, Shanghai Institute of Microsystem and Information Technology, Chinese Academy of Sciences, Shanghai 200050, China}
\author{Yang Liu}
\affiliation{Department of Modern Physics and National Laboratory for Physical Sciences at Microscale, Shanghai Branch, University of Science and Technology of China, Hefei, Anhui 230026, China}
\affiliation{CAS Center for Excellence and Synergetic Innovation Center in Quantum Information and Quantum Physics, Shanghai Branch,  University of Science and Technology of China, Hefei, Anhui 230026, China}
\author{Wei-Jun Zhang}
\affiliation{State Key Laboratory of Functional Materials for Informatics, Shanghai Institute of Microsystem and Information Technology, Chinese Academy of Sciences, Shanghai 200050, China}
\author{Xiao Jiang}
\affiliation{Department of Modern Physics and National Laboratory for Physical Sciences at Microscale, Shanghai Branch, University of Science and Technology of China, Hefei, Anhui 230026, China}
\affiliation{CAS Center for Excellence and Synergetic Innovation Center in Quantum Information and Quantum Physics, Shanghai Branch,  University of Science and Technology of China, Hefei, Anhui 230026, China}
\author{Lu Zhang}
\affiliation{State Key Laboratory of Functional Materials for Informatics, Shanghai Institute of Microsystem and Information Technology, Chinese Academy of Sciences, Shanghai 200050, China}
\author{Jian Wang}
\affiliation{Department of Modern Physics and National Laboratory for Physical Sciences at Microscale, Shanghai Branch, University of Science and Technology of China, Hefei, Anhui 230026, China}
\affiliation{CAS Center for Excellence and Synergetic Innovation Center in Quantum Information and Quantum Physics, Shanghai Branch,  University of Science and Technology of China, Hefei, Anhui 230026, China}
\author{Li-Xing You}
\affiliation{State Key Laboratory of Functional Materials for Informatics, Shanghai Institute of Microsystem and Information Technology, Chinese Academy of Sciences, Shanghai 200050, China}
\author{Jian-Yu Guan}
\author{Dong-Xu Yang}
\affiliation{Department of Modern Physics and National Laboratory for Physical Sciences at Microscale, Shanghai Branch, University of Science and Technology of China, Hefei, Anhui 230026, China}
\affiliation{CAS Center for Excellence and Synergetic Innovation Center in Quantum Information and Quantum Physics, Shanghai Branch,  University of Science and Technology of China, Hefei, Anhui 230026, China}
\author{Zhen Wang}
\affiliation{State Key Laboratory of Functional Materials for Informatics, Shanghai Institute of Microsystem and Information Technology, Chinese Academy of Sciences, Shanghai 200050, China}
\author{Hao Liang}
\affiliation{Department of Modern Physics and National Laboratory for Physical Sciences at Microscale, Shanghai Branch, University of Science and Technology of China, Hefei, Anhui 230026, China}
\affiliation{CAS Center for Excellence and Synergetic Innovation Center in Quantum Information and Quantum Physics, Shanghai Branch,  University of Science and Technology of China, Hefei, Anhui 230026, China}
\author{Zhen Zhang}
\affiliation{Center for Quantum Information, Institute for Interdisciplinary Information Sciences, Tsinghua University, Beijing, 100084, China}
\affiliation{CAS Center for Excellence and Synergetic Innovation Center in Quantum Information and Quantum Physics, Shanghai Branch,  University of Science and Technology of China, Hefei, Anhui 230026, China}
\author{Nan Zhou}
\affiliation{Department of Modern Physics and National Laboratory for Physical Sciences at Microscale, Shanghai Branch, University of Science and Technology of China, Hefei, Anhui 230026, China}
\affiliation{CAS Center for Excellence and Synergetic Innovation Center in Quantum Information and Quantum Physics, Shanghai Branch,  University of Science and Technology of China, Hefei, Anhui 230026, China}
\author{Xiongfeng Ma}
\affiliation{Center for Quantum Information, Institute for Interdisciplinary Information Sciences, Tsinghua University, Beijing, 100084, China}
\affiliation{CAS Center for Excellence and Synergetic Innovation Center in Quantum Information and Quantum Physics, Shanghai Branch,  University of Science and Technology of China, Hefei, Anhui 230026, China}
\author{Teng-Yun Chen}
\author{Qiang Zhang}
\author{Jian-Wei Pan}
\affiliation{Department of Modern Physics and National Laboratory for Physical Sciences at Microscale, Shanghai Branch, University of Science and Technology of China, Hefei, Anhui 230026, China}
\affiliation{CAS Center for Excellence and Synergetic Innovation Center in Quantum Information and Quantum Physics, Shanghai Branch, University of Science and Technology of China, Hefei, Anhui 230026, China}

\begin{abstract}
A main type of obstacles of practical applications of quantum key distribution (QKD) network is various attacks on detection. Measurement-device-independent QKD (MDIQKD) protocol is immune to all these attacks and thus a strong candidate for network security.
Recently, several proof-of-principle demonstrations of MDIQKD have been performed. Although novel, those experiments are implemented in the laboratory with secure key rates less than 0.1 bps. Besides, they need manual calibration frequently to maintain the system performance. These aspects render these demonstrations far from practicability.
Thus, justification is extremely crucial for practical deployment into the field environment.
Here, by developing an automatic feedback MDIQKD system operated at a high clock rate, we perform a field test via deployed fiber network of 30 km total length,
achieving a 16.9 bps secure key rate. The result lays the foundation for a global quantum network which can shield from all the detection-side attacks.
\end{abstract}
\maketitle

\maketitle

\section{Introduction}
 Quantum key distribution (QKD) can in principle provide information-theoretical security based on quantum mechanics. It is the most practical application of the fast developing field of quantum information technology. After some early experimental demonstrations carried out in 1990s to verify the feasibility of QKD, in 2000s the QKD systems are successfully transformed from controlled laboratory environments to real-life implementations \cite{TYChen:fieldtest:2009,Peev:SECOQC:2009,Sasaki:TokyoQKD:2011}, to realize the practical value of QKD.  Up till now, quite a few commercial QKD systems are available in the market \cite{CommercialQKD} and are under rapid development.

Despite the progress considering either the experiment development or commercialization, practical QKD systems are suffering from various attacks that render the QKD systems insecure \cite{MAS_Eff_06,Fung:Remap:07,Qi:TimeShift:2007,Zhao:TimeshiftExp:2008,Lydersen:Hacking:2010,Weier:DeadtimeAttack:2011,Mark:DamageAtt:2014}. These attacks take advantage of the imperfections, especially the detection-side ones, rooted in the gap between the theoretical model and the practical QKD systems.
%In order to achieve information-theoretical security when imperfect devices are present, device independent QKD schemes \cite{Acin:DeviceIn:07} (without assumptions on either the detector or the source) have been proposed. However, their unrealistic requirement for a high transmission efficiency together with an extremely low secure key rate, limits their practical usage.
Some of them are experimentally demonstrated \cite{Zhao:TimeshiftExp:2008,Lydersen:Hacking:2010} based on these practical QKD systems.
Although certain countermeasures are provided to close some specific side channels \cite{Scarani:QKD:2009,Zhao:TimeshiftExp:2008,Yuan:AvoidingAttack:2010}, there might still be some side channels which are hard to estimate and will cause potential threats.
%\cite{MayersYao_98,Acin:DeviceIn:07}
A practical QKD system that can close all the detection-side loopholes is still missing.

Recently, the newly proposed measurement-device-independent QKD (MDIQKD) \cite{Lo:MIQKD:2012} protocol, whose security does not rely on any assumption on the detection system, can defeat all the detection-side attacks. Many efforts have been made extensively on the experimental demonstrations of the MDIQKD protocol \cite{Tittel:MDIQKDFielfTest:2013,Liu:MIQKDexp:2013,Silva:DemoPolMDIQKD:2013,tang:experimental:2013}. These results they have achieved demonstrate the feasibility of MDIQKD, while they are far away from practicability. Generally, there are three basic criteria for a practical QKD system: stabilization under real-world environment, a moderate secure key rate, and an automatic operation.

Firstly, all previous demonstrations are taken in the laboratory without perturbation of the field environment. An field test \cite{Tittel:MDIQKDFielfTest:2013} of the MDIQKD scheme has been attempted over an 18.6 km deployed fiber (9 dB transmission loss), but have not generated secure key actually, since the decoy-state method is not adopted in this experiment. Secondly, the secure key rates in all previous experiments \cite{Tittel:MDIQKDFielfTest:2013,Liu:MIQKDexp:2013,Silva:DemoPolMDIQKD:2013,tang:experimental:2013} are limited, of which the highest is $0.12\ bps$ at 50 km transmission distance (10 dB transmission loss) \cite{Liu:MIQKDexp:2013}. Last but not least, all previous experiments need manual calibration frequently to maintain the system performance per 10 minutes. This is fatal for a practical application. A sufficiently good performance will involve many aspects, such as time, spectrum and polarization modes. This poses another challenge on implementing an automatic calibration system. It is wondered whether the MDIQKD system is suitable for a practical deployment or not.

\section{Experimental Setup}
\begin{figure}[tbh]
\centering
\resizebox{8.6cm}{!}{\includegraphics{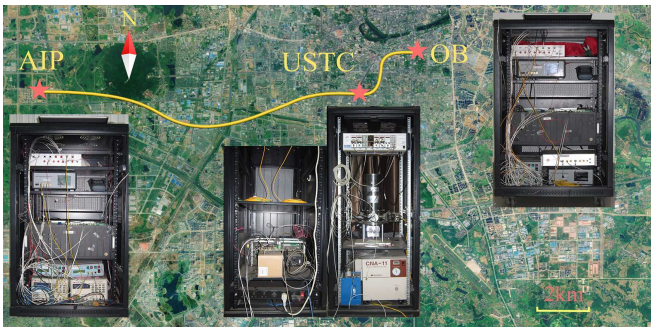}}
\caption{Bird's-eye view of the field-environment MDIQKD. Alice is placed in Animation Industry Park in Hefei (AIP), Bob in an office building (OB), and Charlie in the University of Science and Technology of China (USTC). Alice (Bob) is on the west (east) side of Charlie. AIP-USTC link is 25 km (7.9 dB), and OB-USTC link is 5 km (1.3 dB).}
\label{Fig:FieldTest}
\end{figure}

In this work, we take the field test in three adjacent sites located in Hefei City, China.
We adopt the running fiber network of Hefei Cable Television Broadband Network Corp Ltd due to the low dispersion, low attenuation of the optical fiber at the telecom wavelengths.
As shown in Fig.~\ref{Fig:FieldTest}, Alice is placed in the site of Animation Industry Park in Hefei (AIP) $(N31^{\circ}50^{'}6^{''}, E117^{\circ}7^{'}52^{''})$, Bob in the site of an office building(OB) $(N31^{\circ}50^{'}57^{''}, E117^{\circ}16^{'}50^{''})$ and Charlie in the campus of University of Science and Technology of China (USTC) $(N31^{\circ}50^{'}8^{''}, E117^{\circ}15^{'}47^{''})$. The total deployed fiber length is 30 km, with AIP-USTC link of 25 km (7.9 dB) and OB-USTC link of 5 km (1.3 dB). The signal laser pulses are transmitted through the two links.  The auxiliary synchronization laser and the phase-stabilization laser in the feedback systems are multiplexed by the wavelength division multiplexer (WDM), and are transmitted through two additional fiber links.

\begin{figure*}[tbh]
\centering
\resizebox{14cm}{!}{\includegraphics{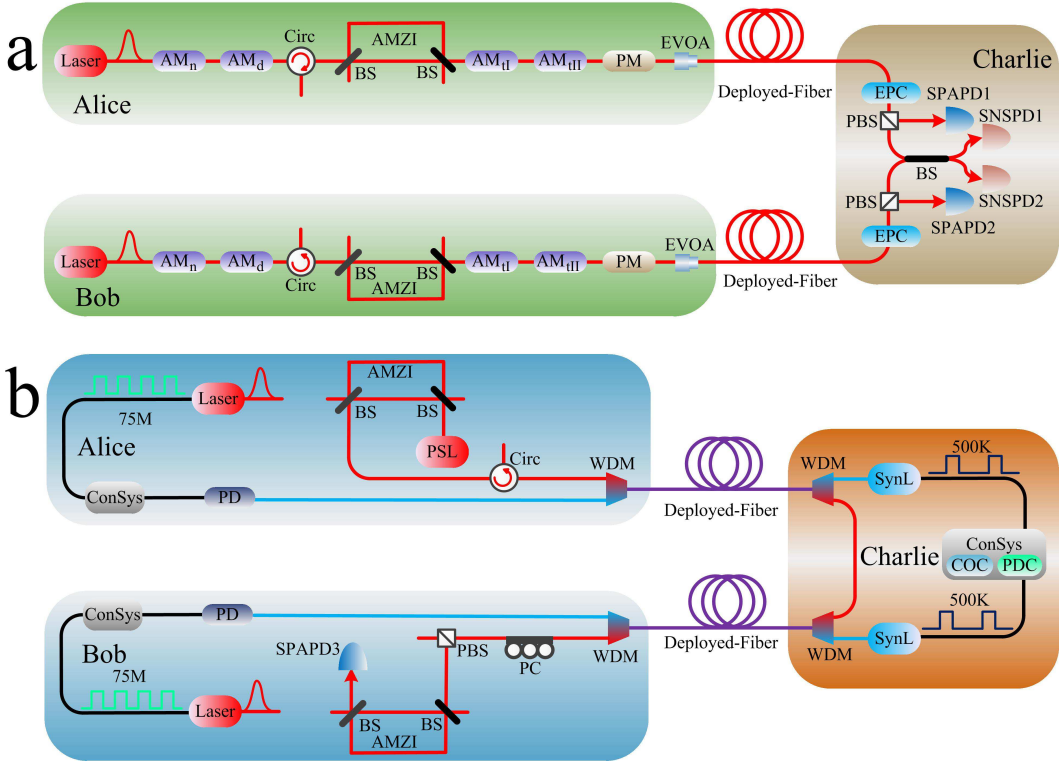}}
\caption{(a) Diagram of our experimental setup. The signal laser source is phase randomized by internal modulation. AM$_\textmd{d}$ is adopted to modulated Alice's (Bob's) signal laser pulses (1550 nm) into three decoy-state intensities.
AMZI, AM$_\textmd{tI}$, AM$_\textmd{tII}$, AM$_\textmd{n}$ and one PM are used to encode qubits. A circulator (Circ) is inserted before AMZI to separate the forward signal laser pulses from the backward PSL pulses.
In the measurement station, Charlie receives the pulses sent through the deployed fiber links, stabilizes the input polarization by the polarization stabilization system (comprised of an EPC, a PBS and a SPAPD), and then takes a partial BSM (implemented with an interference BS and two SNSPDs).
(b) Time calibration system and phase stabilization system. The time calibration system adopts two synchronization lasers (SynLs, 1570 nm), with the 500 kHz shared time reference generated from a crystal oscillator circuit (COC) and with the time delayed by a programmable delay chip (PDC) within the control system (ConSys). The SynLs are transmitted to Alice and Bob through two additional fiber links, respectively. The phase stabilization system utilizes a PSL with the same wavelength as the signal laser's. With the help of WDM, the Alice-Charlie fiber link and the Bob-Charlie fiber link are combined to be the channel transmitting the PSL pulses. PC: polarization controller, PS: phase shifter. }
\label{Fig:LabMDIQKD}
\end{figure*}

In this field-environment test, we develop a decoy-state MDIQKD system, operated at a clock rate of 75 MHz and with a superconducting nanowire single photon detector (SNSPD) system of more than 40\% detection efficiency \cite{Yang:SNSPD:2014}. Our experimental setup is illustrated in Fig.~\ref{Fig:LabMDIQKD}. To rule out the unambiguous-state-discrimination attack \cite{tang:USDAttack:2012}, we have utilized the internally modulated signal laser source which is intrinsically phase randomized. Besides, we employ the vacuum+weak decoy-state scheme \cite{Hwang:Decoy:2003,Lo:Decoy:2005,Wang:Decoy:2005} to defeat the PNS attack \cite{Brassard:PNS:2000}. According to the decoy-state method, Alice (Bob) randomly sets the laser pulse intensity to be among three different values, ${0,\nu=0.07,\mu=0.40}$, as the intensities of vacuum state, weak decoy state and signal state. Their probabilities are set as $22\%$, $45\%$ and $33\%$, respectively.
%The decoy state intensity and probability distribution are optimized to maximize the secure key rate.
We employ the time-bin phase-encoding scheme \cite{Lo:MIQKD:2012,MXF:MIQKD:2012}, and utilize an asymmetrical Mach-Zehnder interferometer (AMZI), three AMs and one PM to encode qubits.
AMZI splits the laser pulse into two time bins with a 6.5 ns time delay.
If $Z$ basis is used, the key bit is encoded in only one time bin by two AMs. If $X$ basis is used, the key bit is encoded into two time bins' relative phase, 0 or $\pi$, by PM.
The random basis and bit choices are of uniform probabilities.
Another AM in the three AMs serves to normalize the average photon numbers in the two bases.
The electrical variable optical attenuator (EVOA) is to attenuate the laser's output intensity to single photon level.
We remark that two AMs are employed to not only increase the fidelity of time bin 0 or 1, but also improve the extinction ratio of the vacuum state intensity for the decoy-state method.

The laser pulses of Alice (Bob) go through the Alice-Charlie (Bob-Charlie) fiber link, to interfere with the ones sent by Bob (Alice). Charlie in the measurement station then takes a partial Bell-state measurement (BSM) implemented with an interference beam splitter (BS) and two SNSPDs at the two output arms of the BS. Then Charlie announces the BSM results to Alice and Bob for them to distill the secure key.
Bell state $| \psi^- \rangle$ is post-selected when the two SNSPDs have a coincidence detection at two alternative time bins, i.e., SNSPD1 has a detection at time bin 0 (1) and simultaneously SNSPD2 has a detection at time bin 1 (0). The information of Alice and Bob are thus anti-correlated. Alice just needs to flip all the key bits to get correlated key stream with Bob's.

\section{Automatical Feedback Systems}
In order to achieve both a highly efficient coincidence count rate and a desirable error rate, we require a perfect and stable BSM, namely, the two independent laser pulses should keep indistinguishable after traveling through two separated fiber links, especially in the scenario of an unstable field environment.
Thus, three aspects, time, spectrum and polarization, should be taken into account. To maintain the system performance and continuous operation, we develop several automatical feedback systems, serving for calibrating the time, spectrum and polarization modes of two independent laser pulses.

For the time synchronization of the whole system shown in Fig.~\ref{Fig:LabMDIQKD}(b), two synchronization laser (SynL) pulse trains are directly modulated by 500 KHz electric signals from a crystal oscillator circuit, and are sent from Charlie to Alice and Bob, respectively. Alice (Bob) utilizes a photoelectric detector (PD) to detect the SynL pulses. The output signals of the PD are used to regenerate a 75 MHz system clock as the time reference for the signal lasers and all amplitude and phase modulators. Thus the whole system becomes synchronized.

Then we precisely overlap the two signal pulses through a feedback control. Alice and Bob alternatively send her (his) signal laser pulses to Charlie.  She (He) increases the intensity of the output signal laser pulses by adjusting the EVOA, so that Charlie can get enough detection events of SNSPD to calculate the average arriving time of Alice's (Bob's) signal laser pulses within several seconds. Based on the arriving time difference, Charlie adjusts the time delay between the two SynL pulses with a programmable delay chip.

There are several aspects that can influence the system's timing jitter: 1) the programmable delay chip that adjusts the time delay
of the SynL's triggering signal. As in our experiment, the timing jitter increases with the time delay value. 2) the received power of the SynL pulses and the distinguishing voltage level of the electronics circuit with PD on it. They both should be optimized correspondingly at a certain transmission distance. 3) the SNSPD used in the BSM module. Besides the merits of high detection efficiency and low dark counts, the SNSPD has another advantage of low timing jitter within 100 ps that largely improves the overall timing jitter performance \cite{Chen:SNSPD:2013,You:SNSPD:2013}. 4) the time interval analyzer, which in our experiment is a high-performance time-to-digital converter (TDC). It records the time between the input detection event and its start signal. The start signal here is of the same clock rate with the SynL pulse. Since the timing jitter of TDC gets better with a smaller measurement range of recorded time, we set the SynL clock rate to be 500 kHz in our setup. Thus, considering all the aspects $1)\sim5)$, as well as that of the time calibration system, we can confirm a good pulse overlap of the time mode.

For the spectrum mode, we firstly select two nearly identical laser diodes as Alice's and Bob's laser sources, considering the aspects of both the same full width at half maximum (FWHM) wavelength and the same central wavelength. Then, we develop a temperature controlled circuit in the laser to automatically stabilize the wavelength. At last, to calibrate the wavelength precisely, we utilize an optical spectrum analyzer to measure the central wavelengths of Alice's and Bob's lasers alternatively.  The temperatures are readjusted through temperature controlled circuits based on the wavelength difference. The temperature controlling precision is 0.005 degree centigrade, and the wavelength controlling precision is about 0.5 pm. Then the difference between Alice's and Bob's spectrum modes can be controlled to an acceptably low level compared with 16 pm (FWHM).

For the polarization mode, we insert an electric polarization controller (EPC) and a polarization beam splitter (PBS) before the interference BS, and connect the transmission port of PBS with the BS. The reflection port of the PBS is monitored by an InGaAs/InP single-photon avalanche photodiode (SPAPD), whose count rate is used as the feedback signal to control the EPC, to guarantee that the maximum amount of laser power is transmitted through the PBS. Using this polarization stabilization system, we can control and eliminate the polarization change to an acceptable low level.

For the phase-encoding scheme of MDIQKD, an important task is to make sure Alice and Bob use the same phase reference frame to avoid the $X$-basis misalignment. The phase reference frame, namely the relative phase between AMZI's two arms, may fluctuate with temperature and stress. Thus, we firstly put the AMZIs in a thermal container to isolate it from the temperature and stress perturbation. Besides, we adopt a phase stabilization system \cite{Liu:MIQKDexp:2013,Tang:MDIQKD200km:2014} to maintain the phase reference frames of Alice's and Bob's AMZIs, as shown in Fig.~\ref{Fig:LabMDIQKD}(b).
We employ a phase-stabilization laser (PSL) with the same wavelength as the signal laser's. The PSL pulses are sent through Alice's and Bob's AMZIs connected by an auxiliary fiber link, and are monitored by another InGaAs/InP SPAPD at an output of Bob's AMZI.
The phase difference is then calibrated by a phase shifter inside Bob's AMZI. Note that the auxiliary fiber link between Alice and Bob is comprised of the auxiliary Alice-Charlie link and Bob-Charlie link, and the PSL pulses are multiplexed with SynL pulses by WDM.

All the aforementioned feedback systems contribute to a good interference and a minimal basis misalignment, and the automatic calibration procedure can largely improve the time utilization efficiency. While the polarization and phase stabilization systems are operated in real time, the time and spectrum calibration systems need to operate in a calibration procedure alternative with the QKD process. We switch the QKD process to calibration procedure every half an hour.

\section{Experimental Results and Secure key Calculation}
Using this MDIQKD system integrated with the feedback systems, we has accumulated the raw data for 18.2 hours. During this period, the feedback systems work effectively. Compared with all the previous laboratory experiments, the field test faces much more severe environment turbulence. Especially, the polarization mode dispersion of the deployed fiber may cause the polarization overlap of the two independent lasers fluctuating. With the help of the aforementioned polarization stabilization system, we have compensated for the polarization change and achieved a fluctuation less than 3\%. Besides, the field environment will also change the arriving time of the signals. The time calibration system works to monitor the time shift and then compensate it effectively. The achieved timing calibration precision is below 20 ps, which is much smaller than the 2.5 ns pulse width of the signal laser.

Furthermore, monitored by the optical spectrum analyzer, the wavelength difference between Alice's and Bob's signal laser sources can be maintained within 1 pm for a few hours. It indicates that the temperature controlled circuit built in the signal laser source can control the wavelength of the laser precisely and stabilize it effectively. Thanks to the feedback procedure with these aforementioned feedback systems, as well as the temperature controlled circuit, this MDIQKD system can run continuously for a long time without manual efforts.

We adopt in the BSM a time window of 1.5 ns, 60\% of the pulse width of 2.5 ns. The details for the experimental results, including the coincidence event counts $M$ and quantum bit error rates (QBERs) $E$, can be found in Table \ref{Tab:DataDetailGain} and \ref{Tab:DataDetailQBER}. Here, $M^{\mu_a\mu_b}$ ($E^{\mu_a\mu_b}$) is the overall coincidence event count (error rate) given that Alice sends out her state with an intensity of $\mu_a$ and Bob sends out his state with an intensity of $\mu_b$. In the $Z$ basis, the results show both a high-efficient coincidence count rate and a desirable low error rate (the QBERs are less than $0.1\%$ when the intensities are neither vacuum). In the $X$ basis, note that $M_{x}^{\mu\mu} \approx M_{x}^{\nu\mu} > M_{x}^{\mu\nu}$, because when Alice's and Bob's pulses interfere in Charlie's site with quite different intensities, the coincidence event count is mainly determined by the larger intensity. Since the transmission loss of Alice-Charlie link is about 6 dB higher than that of the Bob-Charlie link, Alice's laser pulses contributes to the coincidence event count much less than Bob's. Besides, among all the QBERs in the $X$ basis (when the intensities are neither vacuum), the minimal error rate is $E_{x}^{\mu\nu}$, not $E_{x}^{\mu\mu}$, since the QBER in the $X$ basis gets better when the two received pulses interfere with each other with less different intensities.

\begin{table}
\centering
\caption{List of the total coincidence event counts of Bell state $| \psi^- \rangle$ in the 30 km field test for 18.2 hours.} \label{Tab:DataDetailGain}
\begin{tabular}{c||c|ccc}
\hline
\hline
$$ & $\mu_a/\mu_b$ & $0$&$\nu$&$\mu$ \\
\hline
$$                  &  $0$ &$0.00\times10^{0}$	&$1.93\times10^{2}$&$2.64\times10^{3}$   \\
$M_{z}^{\mu_a\mu_b}$&$\nu$ &$3.60\times10^{1}$	&$8.12\times10^{5}$&$3.36\times10^{6}$  \\
$$                  &$\mu$ &$1.46\times10^{2}$	&$3.49\times10^{6}$&$1.35\times10^{7}$ \\
\hline
$$                  &  $0$ &$0.00\times10^{0}$	 &$8.58\times10^{5}$ &$2.03\times10^{7}$  \\
$M_{x}^{\mu_a\mu_b}$&$\nu$ &$4.30\times10^{4}$	 &$2.72\times10^{6}$ &$4.42\times10^{7}$   \\
$$                  &$\mu$ &$9.94\times10^{5}$	 &$6.55\times10^{6}$ &$4.48\times10^{7}$  \\
\hline
\hline
\end{tabular}
\end{table}

\begin{table}
\centering
\caption{List of the QBERs in the 30 km field test for 18.2 hours.} \label{Tab:DataDetailQBER}
\begin{tabular}{c||c|ccc}
\hline
\hline
$$ & $\mu_a/\mu_b$ & $0$&$\nu$&$\mu$ \\
\hline
$$                   & $0$ &  $0.00\% $& $52.33\%$&$49.26\%$ \\
$E_{z}^{\mu_a\mu_b}$&$\nu$ &  $52.78\%$& $0.04\%$&$0.10\%$ \\
$$                  &$\mu$ &  $47.26\%$& $0.01\%$&$0.02\%$ \\
\hline
$$                   & $0$ &  $0.00\%$& $51.49\%$&   $49.90\%$ \\
$E_{x}^{\mu_a\mu_b}$&$\nu$ &  $52.10\%$&$38.12\%$&  $46.85\%$ \\
$$                  &$\mu$ &  $49.92\%$&$27.72\%$&  $36.82\%$ \\
\hline
\hline
\end{tabular}
\end{table}

The final key is assumed to be extracted from the case where both Alice and Bob encode the states in the $Z$ basis using signal states in the decoy-state method. The final secure key length is given by \cite{Lo:MIQKD:2012}:
\begin{equation} \label{keyrate}
\begin{aligned}
K^{\mu\mu} &\ge M_{11}^{\mu\mu}[1-H(e_{11}^{p\mu\mu})]-K_{ec}^{\mu\mu}, \\
K_{ec}^{\mu\mu} &= M^{\mu\mu}fH(E^{\mu\mu}), \\
\end{aligned}
\end{equation}
Here, $M_{11}^{\mu\mu}$ and $e_{11}^{p\mu\mu}$, the coincidence event counts and phase error rate when both sources generate single-photon states within signal states, can be estimated through the decoy-state method.
$K_{ec}^{\mu\mu}$ denotes the number of the secure bits cost in error correction, $f$ is the inefficiency of error correction, and $H(x)=-x \log_2 x -(1-x) \log_2 (1-x)$ is the binary Shannon entropy function. %We denote $e_{11}^{p\mu\mu}$ to be the phase error rates in the $M_{11}^{\mu\mu}$ events, and $e_{x11}^{b}$ to be the bit error rate when both Alice and Bob send single photon states in the $X$ basis.

\begin{figure}[tbh]
\centering
\resizebox{8.6cm}{!}{\includegraphics{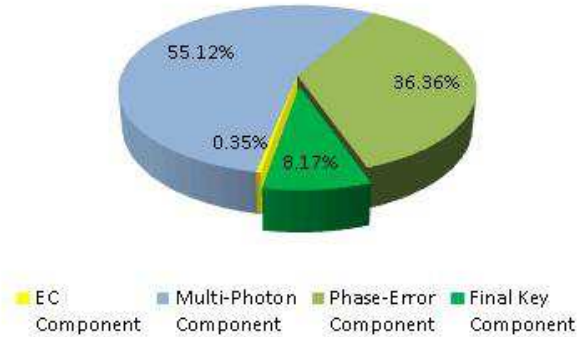}}
\caption{The main results of postprocessing through decoy-state method and the finite-key analysis. Each part denotes its corresponding ratio in the raw key. The final key ratio is $8.17\%$, and the secure key rate is $16.9\ bps$.}
\label{Fig:KeyRatio}
\end{figure}

To calculate the secure key rate from the experimental results, we use the decoy-state method and make the finite-key analysis with the Chernoff bound \cite{curty:finite:2014} to estimate $M_{11}^{\mu\mu}$ and $e_{11}^{p\mu\mu}$, following the same postprocessing detailed in \cite{Tang:MDIQKD200km:2014}. The main results of the postprocessing, as shown in Fig.~\ref{Fig:KeyRatio}, are calculated as follows:
\begin{enumerate}
\item From the experimental data, we can directly obtain $K_{ec}^{\mu\mu}$ = $4.7485\times{10^4}$. This ratio of error correction consumption in the raw key is thus 0.35\%.
\item We can get the lower bound of $M_{11}^{\mu\mu}= 6.0671\times{10^6}$ through the decoy-state method and finite-key analysis. The ratio of multi-photon component ($M^{\mu\mu}-M_{11}^{\mu\mu} = 7.4528\times{10^6}$) is thus 55.12\%.
\item Using the random sampling model \cite{fung:practicalQKD:2010}, we can get the upper bound of $e_{11}^{p\mu\mu}= 24.93\%$. Thus the ratio of the phase error correction part ($M_{11}^{\mu\mu}H(e_{11}^{p\mu\mu}) = 4.9153\times{10^6}$) is 36.36\%.
\item To sum up, the final key length is $1.1046\times{10^6}$, which is 8.17\% of $M^{\mu\mu}$.
\end{enumerate}
Dividing the final key length by the total time in second, we get the secure key rate of $16.9\ bps$.

\section{Conclusion}
The field test demonstrates the feasibility and robustness of the MDIQKD protocol in an unstable environment. In this test, we have generated a final key rate of $16.9\ bps$, which is at least two orders of magnitude higher than the previous results of MDIQKD demonstrations.

Besides, the goal of regular QKD protocols and the MDIQKD protocol is not restricted to point-to-point communication, but is to realize a global quantum network \cite{NNews:Leapout:2014}. The MDIQKD protocol has an intrinsic property which is desirable for constructing quantum network with the star-type structure, since the detection system placed in Charlie's site in the middle node can be shared by all the transmitters. Furthermore, when more transmitters are added in the network, only laser sources and modulators are needed which are much cheaper and smaller than the detection system. While the existing quantum networks are suffering from various attacks, especially the detection-side ones, the MDIQKD protocol will perfectly shield the QKD network from these existing and potential detection-side attacks. We can expect that the MDIQKD network may be built within reach of current technology in the near future.

We remark that there is still much room for us to make MDIQKD system more practicable. Firstly, we can increase the system clock rate by further minimizing the overall timing jitter. Secondly, with the development of the SNSPD technology, the detection efficiency can be further improved \cite{marsili:SSPD93:2013}. Besides, the dark count rate may be effectively reduced to sub-Hertz \cite{Yang:SNSPD:2014}. Last but not least, with the decoy-state parameters and the basis choice optimized \cite{Xu:OptMDIQKD:2014}, we can expect a faster key rate generation to enable some practical applications. We note that this field test utilizes the system based on which we have implemented a long-distance MDIQKD over 200 km  \cite{Tang:MDIQKD200km:2014}.

\section*{Acknowledgment}
The authors would like to thank Xiaoming Xie and Mianheng Jiang for enlightening discussions. This work has been supported by the National Fundamental Research Program (under Grant No. 2011CB921300, 2013CB336800 and 2011CBA00300), the National Natural Science Foundation of China, the Chinese Academy of
Science, the Quantum Communication Technology Co., Ltd., Anhui, and the Shandong Institute of Quantum Science \& Technology Co., Ltd.

%%%%%%%%%%%%%%%%%%%%%%%%%%%%%%%%%%%%%%%%
% choose a style
%\bibliographystyle{ieeetr}
%\bibliographystyle{unsrt}
\bibliographystyle{apsrev}
%%%%%%%%%%%%%%%%%%%%%%%%%%%%%%%%%%%%%%%%

%%%%%%%%%%%%%%%%%%%%%%%%%%%%%%%%%%%%%%%%%%%%%%%%%%%%%%%%%%%%%%%%%%%
% Acknowledgments

\end{document}